\documentclass[amsmath,amssymb,showkeys,superscriptaddress,aps,prd]{revtex4}
\usepackage{graphicx}
\usepackage{pdfpages}
\usepackage[caption=false]{subfig}
\usepackage{multirow}
\usepackage{appendix}
\usepackage{graphicx,epstopdf}
\usepackage{subfig}

\def\xslash{x\!\!\!\slash }

\def\vel{\left|}
\def\ver{\right|}
\usepackage{colortbl}
\usepackage{xcolor}
\usepackage[colorlinks=true, urlcolor=blue, linkcolor=blue, citecolor=blue]{hyperref}

\begin{document}

\title{Magnetic dipole moment of the $Z_{cs}(3985)$ state: diquark-antidiquark and molecular pictures}

\author{U.~\"{O}zdem}%
\email[]{ulasozdem@aydin.edu.tr}
\affiliation{ Health Services Vocational School of Higher Education, Istanbul Aydin University, Sefakoy-Kucukcekmece, 34295 Istanbul, Turkey}
\author{K. Azizi}
\email[]{kazem.azizi@ut.ac.ir (corresponding author)}
\affiliation{Department of Physics, University of Tehran, North Karegar Avenue, Tehran
14395-547, Iran}
\affiliation{Department of Physics, Dogus University, Acibadem-Kadikoy, 34722 
Istanbul, Turkey}

\date{\today}
 
\begin{abstract}

We calculate the magnetic dipole moment  of the newly observed charged hidden-charmed open strange  $ Z_{cs}(3985)^- $ state, recently observed  by  BESIII Collaboration. Based on the information provided by the experiment and  theoretical studies followed  the observation, we assign the quantum numbers $ J^{P} = 1^{+}$ and the quark composition $ c \bar c s\bar u $ to this state and estimate the magnetic dipole moment of this resonance in both the compact diquark-antidiquark and molecular pictures.  We apply the light cone QCD formalism and use the distribution amplitudes of the on-shell photon with different twists.  The obtained results  in both pictures are consistent with each other within the errors.  The magnitude of the magnetic dipole moment  shows that  it is  accessible in the experiment. 

\end{abstract}
\keywords{Magnetic Dipole Moment, Exotic States, Light Cone QCD}

\maketitle

\section{Introduction}
Since 2003, many exotic states with the names of  XYZ  have been experimentally discovered and received especial attention.  The class of charged charmonium like states, mainly denoted by Z,  constitutes one of the main and important categories in tetraquarks. They cannot be  put in the spectrum of the traditional $ \bar {Q} Q $ states by any means because of their nonzero electric charge:  They must be exotic states with a minimum quark content $ Q \bar {Q} q \bar {q'} $, where $ Q $ stands for the heavy $ c $ or $ b $ quark; and $ q $ and  $ q' $ denote the light quarks. 
%
So far, there are eight members in the cluster of the electrically charged exotic resonances: $Z_c(3900)$, $Z_c(4020)$, $Z_1(4050)$,  $Z_c(4200)$, $Z_2(4250)$, $Z_c(4430)$, $Z_b(10610)$, and $Z_b(10650) $ reported in decays into final states contain a pair of light and heavy quarks  \cite{Choi:2007wga,Aaij:2014jqa,Mizuk:2008me,Ablikim:2013mio,Liu:2013dau,Ablikim:2013wzq,Ablikim:2013emm,Chilikin:2014bkk,Wang:2014hta,Collaboration:2011gja}.
Many efforts have been made to determine  the quark-gluon organization of these states, however,  the nature of most of them is still controversial.
As tetraquark states with quark contents $c\bar{c}u\bar{d}$/$b\bar{b}u\bar{d}$,  family of  these charged exotic states were generally studied as diquark-antidiquark and molecular structures. 
Among  these states, $Z_c(3900)$ and   $ Z(4430) $ have been in the focus of much attention (as examples see  Refs. \cite{Agaev:2016dev,Agaev:2017tzv}  and references therein).   The mass difference between $ Z(4430) $  and $Z_c(3900)$ is roughly equal to the mass difference between $\psi(2S)  $ and $ J/\psi $,  the reason that the  $ Z(4430) $ is considered to be the first radial excitation of the $Z_c(3900)$ state \cite{Agaev:2017tzv}. 
For the recent experimental and theoretical progresses on the exotic states see, for instance,  Refs.~\cite{Faccini:2012pj,Esposito:2014rxa,Chen:2016qju,Ali:2017jda,Esposito:2016noz,Olsen:2017bmm,Lebed:2016hpi,Guo:2017jvc,Nielsen:2009uh,Brambilla:2019esw,Liu:2019zoy, Agaev:2020zad, Dong:2021juy}.

Recently, observation of a  charged hidden-charm tetraquark with a strange quark,  in the invariant mass distribution of $D_s^{-}D^{*0}  $ and $ D_s^{*-} D^0 $ was reported by the BESIII Collaboration~\cite{Ablikim:2020hsk}. This new state is called $ Z_{cs}(3985)^- $  (hereafter,  $ Z_{cs} $).
The information provided by the experiment together with the information available from the theory on the mass of this state (see, for instance, \cite{Azizi:2020zyq}),  the spin-parity of this state is assumed to favor $J^P = 1^+$ and its  quark composition is most likely   $ c \bar c s\bar u $. Thus, it is assumed as the strange partner of the famous   $Z_c(3900)$ state. The mass and width of this state are measured  to be $3982.5^{+1.8}_{-2.6}\pm 2.1~\mbox{MeV}$ and $12.8^{+5.3}_{-4.4}\pm 3.0~\mbox{MeV}$, respectively.
After the experimental observation of this state, many theoretical studies were carried out in which spectroscopic parameters of this new particle were examined \cite{Azizi:2020zyq,Liu:2020nge,Meng:2020ihj,Wang:2020kej,Chen:2020yvq,Cao:2020cfx,Du:2020vwb,Sun:2020hjw,Wang:2020rcx,Wang:2020htx,Wang:2020iqt,Jin:2020yjn,Simonov:2020ozp,Sungu:2020zvk,Ikeno:2021ptx,Guo:2020vmu,Zhu:2021vtd, Wang:2020dgr}. 

 The electromagnetic properties of hadrons, beside their spectroscopic parameters, can help us in determination of their exact nature, substructure and  quantum numbers. These parameters can help us to get useful information on the charge and magnetization distributions as well as their geometric shape. 
 Recall that the  multipole moments of particles  like their dipoles, quadrupoles and octupoles  contain information of the spatial distributions of the charge and magnetization inside the particles and  they are directly related to the spatial distributions of quarks and gluons in hadrons. The values of these observables determine whether the  charge distribution inside the particle is spherical or not and give information about the geometric shape of the particle  whether it is  spherical, oblate, prolate, etc. In the present study, we calculate the electromagnetic form factors of  $ Z_{cs}$  state using the light cone QCD sum rule (LCSR) formalism both in the compact diquark-antidiquark and molecular pictures. By using those form factors at static limit, we extract the magnetic dipole moment of the resonance under study. We use the distribution amplitudes (DAs) of the on-shell real photon state at light cone, which are available in terms of different twists.

 Although the short lifetimes of the $Z_{cs}$ state make the magnetic dipole moment problematic to be measured at present, more data accumulation in different experiments in the future may make this feasible.
 $\Delta^+(1232)$ particle has also a very short lifetime, however, its magnetic dipole moment is extracted from the experimental data on the $\gamma N \rightarrow  \Delta \rightarrow  \Delta \gamma \rightarrow \pi N \gamma $ process \cite{Pascalutsa:2004je, Pascalutsa:2005vq, Pascalutsa:2007wb}.
Hence, one procedure for specification of the electromagnetic multipole moments is relied on soft photon emission off the hadrons suggested in Ref.~\cite{Zakharov:1968fb}. The photon carries information about the higher multipole moments of the particle emitted from, as well. The matrix element for the radiative process can be written with respect to the photon's energy $E_\gamma$ as 
\begin{align}
 M \sim A/E_\gamma + B(E_\gamma )^0 + C E_\gamma +...
\end{align}
The electric charge contribute to the amplitude at order $(E_\gamma )^{-1}$ and the contribution coming from the magnetic dipole moment is characterized by the term $(E_\gamma)^0$. Thus, by measuring the cross section or decay width of the radiative process and ignoring from the small contributions of terms linear/higher order  in $E_\gamma$, one can identify the magnetic dipole moment of the particle under investigation.
 %
 %
 Note that we have previously calculated the magnetic dipole and electric quadrupole of the   $Z_c(3900)$ state in Ref.  \cite{Ozdem:2017jqh} in the same framework. 
 The magnetic moment of the $Z_{cs}$ state has been investigated in the framework of the QCD sum rule and its extension in the weak electromagnetic field by using a molecular type interpolating current in Ref.  \cite{,Xu:2020evn}.
 
 The paper is organized as follows. The next section is devoted to the extraction of the  LCSRs for the magnetic dipole moment in both the pictures. In section III, we numerically analyze the magnetic dipole moment of the $ Z_{cs} $ state. The summary of the results and conclusions are given in the last section.

 \section{Formalism}

To calculate the magnetic  dipole moment of the $Z_{cs}$ state
 within the LCSR, we start with the correlation function (CF)
\begin{equation}
 \label{edmn01}
\Pi _{\mu \nu }(p,q)=i\int d^{4}xe^{ip\cdot x}\langle 0|\mathcal{T}\{J_{\mu}^{Z_{cs}}(x)
J_{\nu }^{Z_{cs}\dagger }(0)\}|0\rangle_{\gamma}, 
\end{equation}%
where  the subindex $\gamma$ stands for the background electromagnetic field, $J_{\mu}(x)$ is the interpolating current of the $Z_{cs}$ state and $\mathcal{T}$ is the time ordering operator.  With the quantum numbers $ J^{P} = 1^{+}$ and the quark content  $ c \bar c s\bar u $,  the $Z_{cs}$ state can be interpolated in the diquark-antidiquark and molecular pictures as 

\begin{eqnarray}
J_{\mu }^{Z_{cs}-Di}(x) &=&\frac{i\epsilon \tilde{\epsilon}}{\sqrt{2}}\Big\{ %
\big[ s_{a}^{T}(x)C\gamma _{5}c_{b}(x)\big] \big[ \overline{u}%
_{d}(x)\gamma _{\mu }C\overline{c}_{e}^{T}(x)\big] -\big[ s_{a}^{T}(x)C\gamma _{\mu }c_{b}(x)\big] \big[ \overline{%
u}_{d}(x)\gamma _{5}C\overline{c}_{e}^{T}(x)\big] \Big\},\nonumber\\
\label{eq:Curr}
 J_{\mu}^{Z_{cs}-Mol}(x)&=&\frac{1}{\sqrt{2}}\Big\{[ \bar c_a(x) i\gamma_5 s_a(x)][\bar u_b(x) \gamma_\mu c_b(x)]
  +[\bar c_a(x) \gamma_\mu s_a(x)][\bar u_b(x) i\gamma_5 c_b(x)]\Big\}.
\end{eqnarray}
where $\epsilon =\epsilon _{abc}$, $\tilde{\epsilon}=\epsilon _{dec}$, $C$ is 
the charge conjugation matrix and $a,b,c,d,e$
are color indices. The currents above carry the same quantum numbers and quark contents, however as is seen,  the four quark in the diquark-type current are made colorless all together  by the help of Levi-Civita symbols while the molecule current consists of two already colorless parts multiplied to each other. They create the compact tetraquark and interacting two-meson molecule  of the same quark content from the vacuum, respectively.
We should also remark that the molecular-type current and, upon Fierz transformations, the diquark-type current  not only couple to  exotic tetraquarks but also  to states of two non-interacting non-exotic mesons, lying below the mass of the multiquark system (see for instance  \cite{Weinberg:2013cfa,Lucha:2021mwx,Kondo:2004cr,Lee:2004xk} and references therein). These contributions can be either subtracted
from the  obtained sum rules or they are  included into parameters of the pole term. For  tetraquarks, the second approach is preferred and it  is applied in
Refs. \cite{Wang:2015nwa,Agaev:2018vag,Sundu:2018nxt}. It appears that
the two-meson states modifies the pole contribution as
\begin{equation}
\frac{1}{m^{2}-p^{2}}\rightarrow \frac{1}{m^{2}-p^{2}-i\sqrt{p^{2}}\Gamma (p)%
},  \label{eq:Modif}
\end{equation}%
\textbf{w}here $\Gamma (p)$\ is the finite width of the tetraquark generated by two-meson scattering states.
 When these effects are properly taken into account in the mass sum rules,  they rescale the
current coupling (residue)  of the tetraquark under consideration leaving its mass unchanged. 
Detailed analyses prove that two-meson scattering contributions are small even for tetraquarks with a large width (see Refs. \cite
{Lee:2004xk,Wang:2015nwa,Agaev:2018vag,Sundu:2018nxt}). 
The effects of two-meson scattering states on   $Z_{c}$-like states including  $Z_{cs}$ have been studied in Refs. \cite{Albuquerque:2021tqd,Albuquerque:2020hio}. The obtained results indicate that, contrary to a qualitative large Nc-counting, the two-meson scattering
contributions to the four-quark spectral functions are numerically negligible  (for more information see \cite{Wang:2020iqt,Wang:2019igl,Wang:2020eew}, as well).
As a result, we use the zero-width single-pole approximation in the present study.


In LCSR method, we will acquire the correlation function once with respect to the hadronic parameters like the electromagnetic form factors  and the second with respect to the QCD  parameters and DAs of the on-shell photon. 
By matching  the coefficients of appropriate Lorentz structures  from both sides  and employing the quark-hadron duality assumption we will able to compute the hadronic observables with respect to the QCD degrees of freedom.

 In hadronic representation,  the CF is calculated by its saturation with the intermediate hadronic states. By performing the four-integral over $x$  we get
\begin{align}
\label{edmn04}
\Pi_{\mu\nu}^{Had} (p,q) = {\frac{\langle 0 \mid J_\mu^{Z_{cs}} \mid
Z_{cs}(p) \rangle}{p^2 - m_{Z_{cs}}^2}} \langle Z_{cs}(p) \mid Z_{cs}(p+q) \rangle_\gamma
\frac{\langle Z_{cs}(p+q) \mid {J^\dagger}_\nu^{Z_{cs}} \mid 0 \rangle}{(p+q)^2 - m_{Z_{cs}}^2} + \cdots,
\end{align}
where  q is the photon momentum and dots stand for the contributions coming from the higher states and
continuum. The matrix element
$\langle 0 \mid J_\mu^{Z_{cs}} \mid Z_{cs} \rangle$ is parameterized as
\begin{align}
\label{edmn05}
\langle 0 \mid J_\mu^{Z_{cs}} \mid Z_{cs} \rangle = \lambda_{Z_{cs}} \varepsilon_\mu^\theta\,,
\end{align}
with $\lambda_{Z_{cs}}$ being the residue of the $Z_{cs}$ state.

In the existence of the external electromagnetic background field, 
the matrix element $\langle Z_{cs}(p) \mid  Z_{cs} (p+q)\rangle_\gamma $ can be expressed  in terms of  the Lorentz invariant form factors as follows~\cite{Brodsky:1992px}:
\begin{align}
\label{edmn06}
\langle Z_{cs}(p,\varepsilon^\theta) \mid  Z_{cs} (p+q,\varepsilon^{\delta})\rangle_\gamma
 &= - \varepsilon^\tau (\varepsilon^{\theta})^\alpha
(\varepsilon^{\delta})^\beta
\Bigg[ G_1(Q^2)~ (2p+q)_\tau ~g_{\alpha\beta}  +
G_2(Q^2)~ ( g_{\tau\beta}~ q_\alpha -  g_{\tau\alpha}~ q_\beta) \nonumber\\
&- \frac{1}{2 m_{Z_{cs}}^2} G_3(Q^2)~ (2p+q)_\tau ~q_\alpha q_\beta  \Bigg],
\end{align}
where $\varepsilon^\delta$ and $\varepsilon^{\theta}$ are the 
polarization vectors of the initial and final $Z_{cs}$
states and $\varepsilon^\tau$ is the polarization vector of the photon.  Here, $G_1(Q^2)$, $G_2(Q^2)$ and $G_3(Q^2)$ are invariant form factors,  with  $Q^2=-q^2$. 

Using Eqs. (\ref{edmn04})-(\ref{edmn06}), the correlation function takes the form,
\begin{align}
\label{edmn09}
 \Pi_{\mu\nu}^{Had}(p,q) &=   \frac{\varepsilon_\rho \lambda_{Z_{cs}}^2}{ [m_{Z_{cs}}^2 - (p+q)^2][m_{Z_{cs}}^2 - p^2]}
 \Bigg[ G_2 (Q^2) \Bigg(q_\mu g_{\rho\nu} - q_\nu g_{\rho\mu} -
\frac{p_\nu}{m_{Z_{cs}}^2}  \big(q_\mu p_\rho - \frac{1}{2}
Q^2 g_{\mu\rho}\big) 
 + \nonumber\\
 &  +
\frac{(p+q)_\mu}{m_{Z_{cs}}^2}  \big(q_\nu (p+q)_\rho+ \frac{1}{2}
Q^2 g_{\nu\rho}\big)
-  
\frac{(p+q)_\mu p_\nu p_\rho}{m_{Z_{cs}}^4} \, Q^2
\Bigg)
\nonumber\\
&
+\mbox{other independent structures}\Bigg]\,+\cdots.
\end{align}
The value of  form factor $G_2(Q^2)$ gives us the  magnetic form factor $F_M(Q^2)$ at different $Q^2$ :
\begin{align}
\label{edmn07}
&F_M(Q^2) = G_2(Q^2).
\end{align}
 At static limit,  $Q^2 = 0 $,   $F_M$ is proportional to the
 magnetic dipole moment $\mu_{Z_{cs}}$ for real photon :
\begin{align}
\label{edmn08}
&\mu_{Z_{cs}} = \frac{ e}{2\, m_{Z_{cs}}} \,F_M(Q^2 = 0).
\end{align}

In  QCD side, the CF in Eq. (\ref{edmn01}), is calculated in deep Euclidean region in terms of QCD degrees of freedom as well as the  DAs of the photon. To this end,   we substitute the explicit forms of the  interpolating currents in the CF and  contract the corresponding quark fields with the help of the Wick's theorem. As a result,  we  get
\begin{eqnarray}\label{edmn11}
\Pi_{\mu\nu}^{QCD-Di}(p,q)&=& -\frac{i}{2}\varepsilon_{abc} \varepsilon_{a'b'c'} \varepsilon_{dec} \varepsilon_{d'e'c'}
 \int d^4 x e^{ipx}\nonumber \\
&& \langle 0 |\Big\{Tr\Big[\gamma_5 \tilde{S}_s^{aa'} (x)\gamma_5 S_c^{bb'}(x)\Big] 
 Tr\Big[\gamma_{\mu} \tilde{S}_c^{e'e} (-x)\gamma_{\nu} S_u^{d'd}(-x)\Big] \nonumber\\
&& - Tr\Big[[\gamma_{\mu} \tilde{S}_c^{e'e} (-x)  \gamma_5 S_u^{d'd}(-x)\Big]Tr\Big[\gamma_{\nu} \tilde{S}_s^{aa'} (x)\gamma_5 S_c^{bb'}(x)\Big] \nonumber \\
&&- Tr\Big[\gamma_5 \tilde{S}_s^{aa'} (x)\gamma_{\mu} S_c^{bb'}(x)\Big] Tr\Big[\gamma_5 \tilde{S}_c^{e'e} (-x)\gamma_{\nu} S_u^{d'd}(-x)\Big] \nonumber \\
&&+ Tr\Big[\gamma_{\nu}  \tilde{S}_s^{aa'} (x)\gamma_{\mu} S_c^{bb'}(x)\Big] Tr\Big[\gamma_5 \tilde{S}_c^{e'e} (-x)\gamma_5 S_u^{d'd}(-x)\Big]  \Big\}| 0 \rangle_\gamma,
\end{eqnarray}
in the diquark-antidiquark picture,  and 
\begin{eqnarray}
\label{neweq}
\Pi _{\mu \nu }^{\mathrm{QCD-Mol}}(p,q)&=&-\frac{i}{2}
\int d^{4}xe^{ipx} \langle 0 | \Big\{ \nonumber\\
&& 
\mathrm{Tr}\Big[\gamma _{5}{S}_{s}^{aa^{\prime }}(x)\gamma _{5}S_{c}^{a^{\prime }a}(-x)\Big]
\mathrm{Tr}\Big[\gamma _{\mu }{S}_{c}^{bb^{\prime }}(x)\gamma _{\nu}S_{u}^{b^{\prime }b}(-x)\Big] \notag \\
&&+\mathrm{Tr}\Big[ \gamma _{5 }{S}_{s}^{aa^{\prime}}(x)\gamma _{\nu}S_{c}^{a^{\prime }a}(-x)\Big] 
\mathrm{Tr}\Big[ \gamma_{\mu }{S}_{c}^{bb^{\prime }}(x)\gamma _{5}S_{u}^{b^{\prime }b}(-x)\Big] \notag \\
&&+\mathrm{Tr}\Big[\gamma _{\mu}{S}_{s}^{aa^{\prime }}(x)\gamma _{5 }S_{c}^{a^{\prime}a}(-x)\Big]  
\mathrm{Tr}\Big[ \gamma _{5}{S}_{c}^{bb^{\prime}}(x)\gamma _{\nu }S_{u}^{b^{\prime }b}(-x)\Big] \notag \\
&&+\mathrm{Tr}\Big[\gamma _{\mu }{S}_{s}^{aa^{\prime }}(x)\gamma _{\nu }S_{c}^{a^{\prime}a}(-x)\Big] 
\mathrm{Tr}\Big[\gamma _{5}{S}_{c}^{bb^{\prime }}(x)\gamma_{5}S_{u}^{b^{\prime }b}(-x)\Big]
 \Big\}| 0 \rangle_\gamma,
\end{eqnarray}
in the  molecular picture, where
\begin{equation*}
\widetilde{S}_{c(q)}^{ij}(x)=CS_{c(q)}^{ij\mathrm{T}}(x)C,
\end{equation*}%
with $S_{q(c)}(x)$ being the full light and charm quark propagators.
The relevant propagators are given as~\cite{Balitsky:1987bk}
\begin{align}
\label{edmn12}
S_{q}(x)&=i \frac{{\xslash}}{2\pi ^{2}x^{4}} 
- \frac{\langle \bar qq \rangle }{12} \Big(1-i\frac{m_{q} \xslash}{4}   \Big)
- \frac{ \langle \bar qq \rangle }{192}m_0^2 x^2  \Big(1-i\frac{m_{q} \xslash}{6}   \Big)
-\frac {i g_s }{32 \pi^2 x^2} ~G^{\mu \nu} (x) \Big[\rlap/{x} 
\sigma_{\mu \nu} +  \sigma_{\mu \nu} \rlap/{x}
 \Big],
\end{align}
and
\begin{align}
\label{edmn13}
S_{c}(x)&=\frac{m_{c}^{2}}{4 \pi^{2}} \Bigg[ \frac{K_{1}\Big(m_{c}\sqrt{-x^{2}}\Big) }{\sqrt{-x^{2}}}
+i\frac{{\xslash}~K_{2}\Big( m_{c}\sqrt{-x^{2}}\Big)}
{(\sqrt{-x^{2}})^{2}}\Bigg]
-\frac{g_{s}m_{c}}{16\pi ^{2}} \int_0^1 dv\, G^{\mu \nu }(vx)\Bigg[ \big(\sigma _{\mu \nu }{\xslash}
  +{\xslash}\sigma _{\mu \nu }\big)\nonumber\\
  &\times \frac{K_{1}\Big( m_{c}\sqrt{-x^{2}}\Big) }{\sqrt{-x^{2}}}
+2\sigma_{\mu \nu }K_{0}\Big( m_{c}\sqrt{-x^{2}}\Big)\Bigg].
\end{align}%
where $\langle \bar qq \rangle$ is quark  condensate, $m_0$ is defined through the quark-gluon mixed condensate  $\langle 0 \mid \bar  q\, g_s\, \sigma_{\alpha\beta}\, G^{\alpha\beta}\, q \mid 0 \rangle = m_0^2 \,\langle \bar qq \rangle $, $G^{\mu\nu}$ is the gluon field strength tensor,  $\sigma_{\mu\nu}= \frac{i}{2}[\gamma_\mu, \gamma_\nu]$ and $K_i$'s are modified Bessel functions of the second kind. 

The correlation functions in Eqs. (\ref{edmn11}) and (\ref{neweq})  contain various contributions: the photon can be emitted both perturbatively or non-perturbatively.
In the first case, the photon interacts with one of the light or heavy quarks, perturbatively. In this situation,  
the propagator of the quark interacting with the photon perturbatively is modified via
\begin{align}
\label{free}
S^{free}(x) \rightarrow \int d^4y\, S^{free} (x-y)\,\rlap/{\!A}(y)\, S^{free} (y)\,,
\end{align}
where $S^{free}(x)$ stands for the first term of the light or heavy quark propagator, and the remaining three propagators in Eqs.~(\ref{edmn11}) and (\ref{neweq}) are substituted with the full quark propagators involving the perturbative and the non-perturbative contributions. The full perturbative contribution  is acquired via carrying out the above substitution for the perturbatively interacting quark propagator with the photon and substituting the other propagators via their free parts.

In the second case, one of the light quark propagators in Eqs.~(\ref{edmn11}) and (\ref{neweq}), characterized the photon emission at large distances, is substituted via
\begin{align}
\label{edmn14}
S_{\mu\nu}^{ab}(x) \rightarrow -\frac{1}{4} \big[\bar{q}^a(x) \Gamma_i q^b(x)\big]\big(\Gamma_i\big)_{\mu\nu},
\end{align}
and the other  propagators are substituted with the full quark propagators.
 Here, $\Gamma_i$ are the full set of Dirac matrices. Once 
Eq. (\ref{edmn14}) is inserted into Eqs. (\ref{edmn11}) and (\ref{neweq}) , there appear matrix
elements like $\langle \gamma(q)\vel \bar{q}(x) \Gamma_i q(0) \ver 0\rangle$
and $\langle \gamma(q)\vel \bar{q}(x) \Gamma_i G_{\mu\nu}q(0) \ver 0\rangle$,
representing the non-perturbative contributions. 
These matrix elements can be written with respect to the photon wave functions with definite
twists, whose expressions are presented in Appendix. The QCD representation of the correlation function can be acquired in connection with quark-gluon parameters as well as the DAs of the photon using Eqs.~(\ref{edmn11})-(\ref{edmn14}) and after applying the Fourier transformation to transfer the computations to the momentum space.

Finally, we choose the structure $q_\mu \varepsilon_\nu$  from the both sides and match its coefficients from both the hadronic and QCD sides. In order to suppress the contributions of the higher states and continuum, we apply Borel transformation and continuum subtraction. The procedures are lengthy but standard, we do not present the steps here and refer the reader for instance to Ref. \cite{Azizi:2018duk}. Finally, we get the sum rules for the  magnetic moment  in two pictures as
\begin{align}
 &\mu_{Z_{cs}}^{Di}\,\, \lambda_{Z_{cs}}^2  = e^{\frac{m_{Z_{cs}}^2}{M^2}} \,\, \Pi^{QCD-Di}(M^2,s_0),\\
 &\mu_{Z_{cs}}^{Mol}\,\, \lambda_{Z_{cs}}^2  =e^{\frac{m_{Z_{cs}}^2}{M^2}}\,\, \Pi^{QCD-Mol}(M^2,s_0),
\end{align}
where $M^2$ and $s_0$ are auxiliary parameters stemming from the applications of the Borel transformation and continuum procedures. The $\Pi^{QCD-Di}(M^2,s_0)$ and $\Pi^{QCD-Mol}(M^2,s_0)$ functions are quite lengthy,  explicit expressions of which  are not presented here.

\section{Numerical analysis}

In this section, we numerically analyze the results of calculations for the magnetic dipole moment in two pictures under consideration.
We use $m_u=m_d=0$, $m_s =96^{+8}_{-4}\,\mbox{MeV}$,
$m_c = (1.275\pm 0.025)\,$GeV,   $m_{Z_{cs}}= 3982.5^{+1.8}_{-2.6}\pm 2.1~\mbox{MeV}$,  
$\langle \bar ss\rangle $= $0.8 \langle \bar uu\rangle$ with 
$\langle \bar uu\rangle $=$(-0.24\pm0.01)^3\,$GeV$^3$~\cite{Ioffe:2005ym},  
$m_0^{2} = 0.8 \pm 0.1$~GeV$^2$~\cite{Ioffe:2005ym}, $\langle \frac{\alpha_s}{\pi} G^2 \rangle =(0.012\pm0.004)$ $~\mathrm{GeV}^4 $~\cite{Belyaev:1982cd}, 
$\lambda_{Z_{cs}}^{Di}=(2.15 \pm 0.44)\times 10^{-2}$~GeV$^5$~\cite{Azizi:2020zyq} and $\lambda_{Z_{cs}}^{Mol}=(2.22^{+0.20}_{-0.17} )\times 10^{-2}$~GeV$^5$~\cite{Xu:2020evn}.
The wavefunctions inside the DAs of the photon and all the related parameters  are borrowed from Ref.~\cite{Ball:2002ps} and presented the appendix.

Apart from these input parameters, the magnetic moment of the $ Z_ {cs} $ state also depends on two auxiliary parameters: the Borel mass parameter ($M^2$) and continuum threshold ($s_0$).
The physical observables studied should show  good stabilities  with respect  to the variations of these auxiliary parameters according to the standard prescriptions of the method. In practice, however,  there appear some dependence specially on $s_0$ leading to some uncertainties in the numerical values of the physical quantities under study.
To fix their  working windows, the conditions of weak dependence of the results on the arbitrary variables, convergence of the OPE and pole dominance are applied.

The continuum threshold is not totally arbitrary but it depends on the energy of the first excited state with the same quantum numbers.   Above the threshold, the  excited states and continuum  begin to contribute to the CF. 
We have no information on the first excited state of this channel, experimentally. Hence, considering the standard requirements, we look for a working region for threshold parameter that the results weakly depend on it and this interval leads to a stable Borel window.  Our numerical results show that the working interval,  $(m_{Z_{cs}} + 0.3 ~\mbox{GeV})^2 \leq s_0 \leq(m_{Z_{cs}} + 0.5~ \mbox{GeV})^2$,   satisfies all the requirements and lead to a high pole contribution.  We choose the interval $18.3~\mbox{GeV}^2 \leq s_0 \leq 20.1~\mbox{GeV}^2$ for the continuum threshold, obtained from the above inequality  by replacing the central experimental value for mass.  The numerical calculations depict that the dependence of the results  on this parameter is relatively weak in this interval.

 The  upper bound of  $M^2$ is found demanding the pole dominance.  i.e.,
\begin{eqnarray}
PC=\frac{\Pi^{QCD}(s_0,M^2)}{\Pi^{QCD}(\infty,M^2)}\geq 0.5,
\end{eqnarray}
where $ PC $ stands for the pole contribution.  The lower bound of $M^2$ is acquired requiring that the perturbative contribution exceeds over the non-perturbative one and the series of non-perturbative operators are convergent.  These conditions lead to the Borel window $4.0~\mbox{GeV}^2 \leq M^2 \leq 6.0~\mbox{GeV}^2 $, where the physical quantities under study demonstrate good stabilities with respect to this parameter. 
Our analyses show that  $ PC $ is roughly $ (58-60)\% $ in the average values of the continuum threshold and Borel parameter.  The higher twist contribution is maximally $ 5\% $ at average values of the auxiliary parameters.

 In Fig. 1, we depict the dependence of the magnetic dipole moment of $   Z_{cs} $   on $M^2$ for both pictures  at three fixed values of  the continuum threshold $s_0$.
  As is seen, the magnetic moment of  $   Z_{cs} $ depicts good stability with respect to $M^2$ in its working window in both pictures.  Although    $\mu_{Z_{cs}}$ shows some dependence on   $s_0$,  it remains inside the limits allowed by the method and  constitutes the main parts of the uncertainties. 
  %
 \begin{figure}[htp]
\centering
 \includegraphics[width=0.47\textwidth]{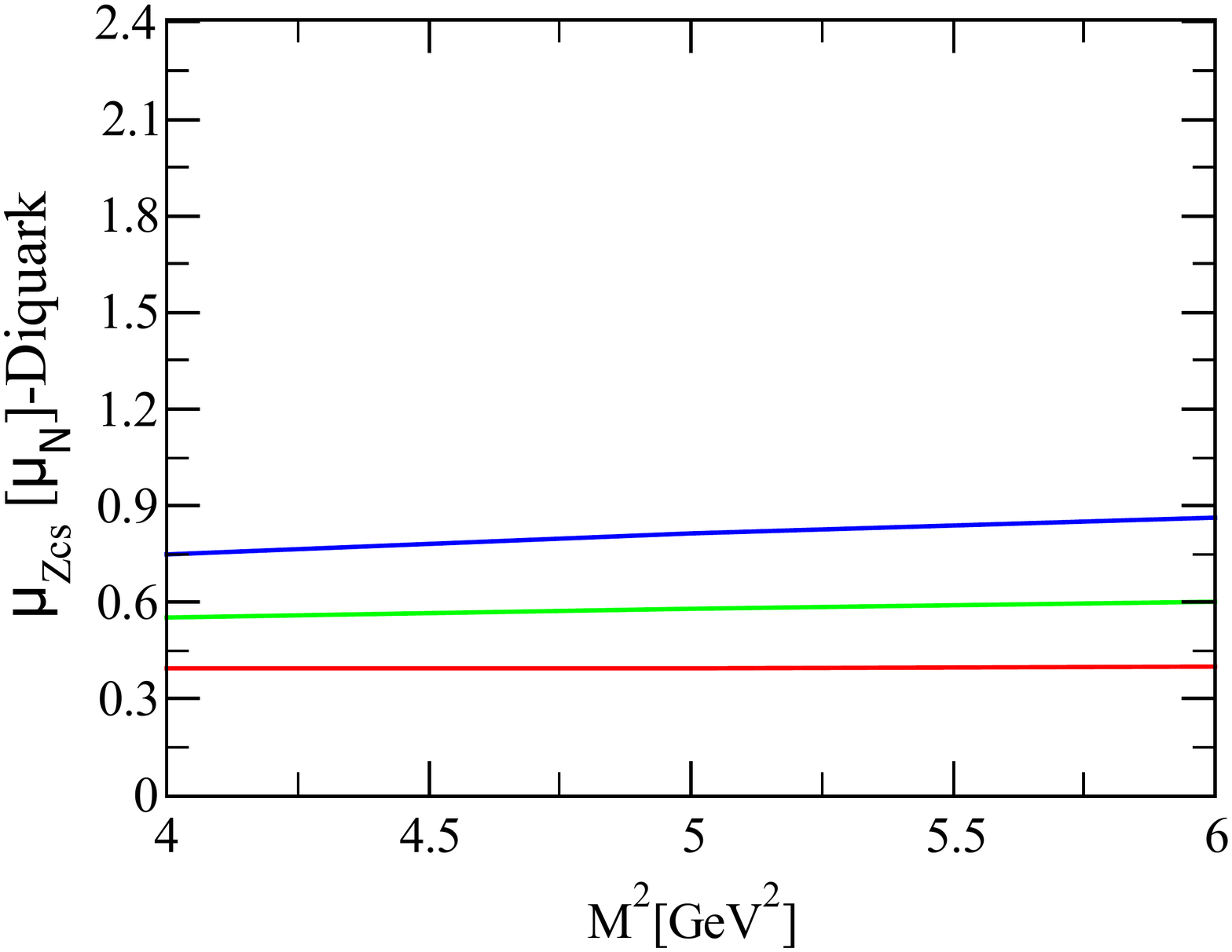}~~
  \includegraphics[width=0.47\textwidth]{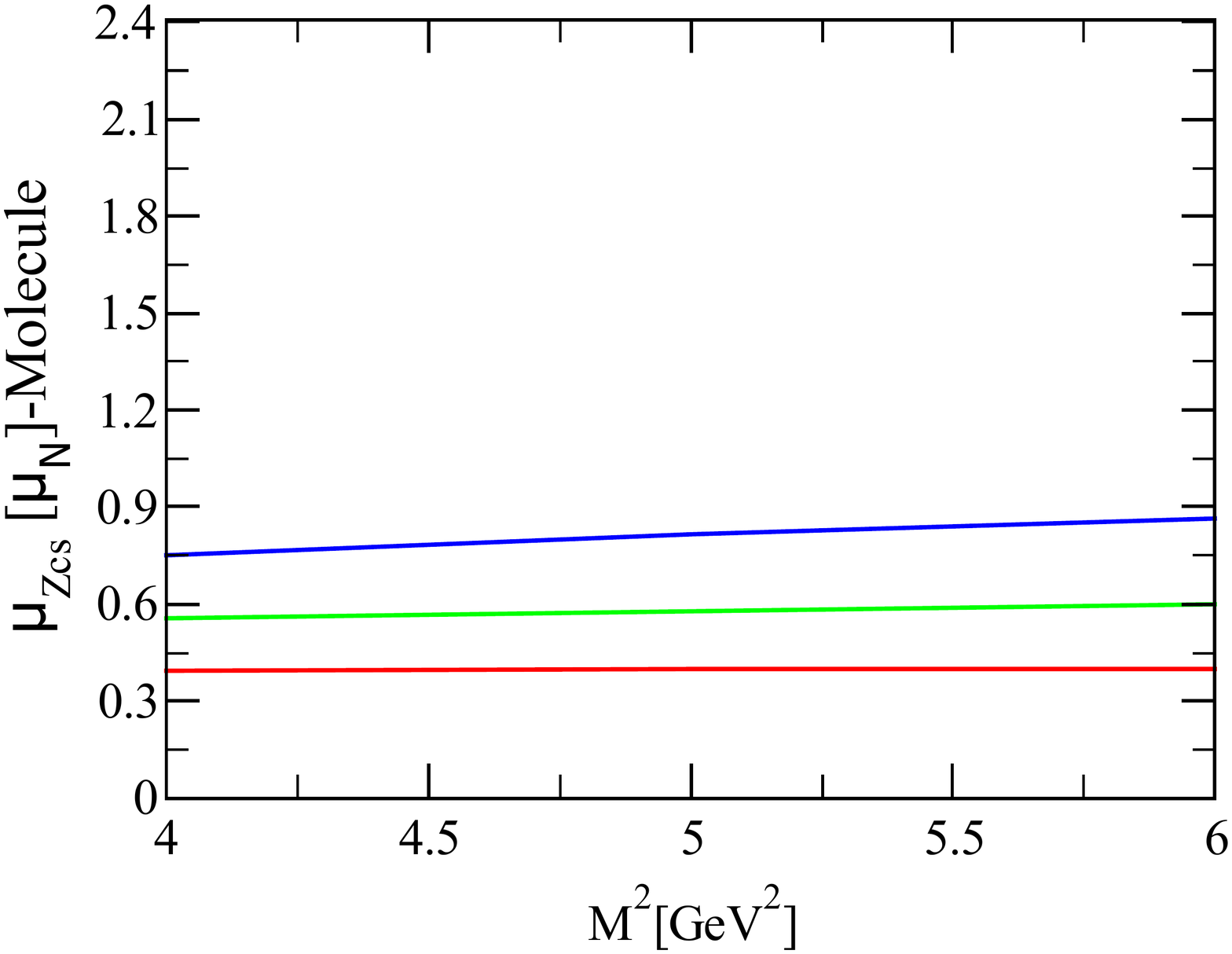}
 \caption{ The dependence of magnetic moment of the $Z_{cs}$ state on $M^{2}$  at three fixed  values of $s_0$ in both pictures. The red, green and blue lines stand for the values  $s_0 = 18.3$ GeV$^2$, $s_0 = 19.2$ GeV$^2$ and $s_0 = 20.1$ GeV$^2$ for the continuum threshold, respectively.}
  \end{figure}

The final  results extracted from the analyses for the magnetic dipole moment of  $Z_{cs}$ state in both the compact tetraquark of diquark-antidiquark and the two-meson molecule pictures are given as
\begin{align}
 &\mu_{Z_{cs}}^{Di} =0.60^{+0.26}_{-0.21}~\mu_N,\nonumber\\
 \mbox{and}\nonumber\\
  &\mu_{Z_{cs}}^{Mol} =0.52^{+0.19}_{-0.17} ~\mu_N,
\end{align}
where the presented errors are related to the uncertainties of the results with respect to the auxiliary parameters and those coming from other inputs. Our results may be checked via other phenomenological methods. 
Comparing the results obtained from both pictures, we see that they are consistent with each other within the presented uncertainties. 
This situation in the tetraquarks is usual as they contain two quarks and two anti-quarks. This is against the situation in the usual baryons made of three quarks: The $   \Sigma$ and $   \Lambda$  as two members of the octet baryons have very different magnetic dipole moments despite the same quark contents \cite{Aliev:2002ra}.  This is attributed to the fact that these baryons have different currents resulting from different diquark-quark structures considering their properties. However, the magnitude of the results on  $\mu_{Z_{cs}}$ indicates that the magnetic dipole moment of the  $Z_{cs}$ is accessible in the experiment. As the results of both pictures on the magnetic dipole moment are roughly equal, comparison of the future experimental data with the theoretical calculations can  not lead us exactly determine the nature and internal quark-gluon organization of this state by comparison of  the obtained results on only the magnetic dipole moment and we need to investigate other parameters of this state as well.  In the literature, there are many attempts to have some  assignments on the substructure of $Z_{cs}$ state.
In Ref. \cite{Azizi:2020zyq},  the mass of the $Z_{cs}$  state was estimated in  compact diquark-antidiquark  picture, which was obtained in a good consistency with  the experimental result. The  mass, current coupling, and vector-self energy of this state  were also investigated in a medium with finite density in this study.
In Ref. \cite{Chen:2020yvq}, the authors have performed a dynamical study on the $D^{(*)0}D_s^{*-}$ interactions by adopting a one-boson-exchange model and taking into account the coupled channel effect. After producing the phase shifts, their results excluded  the newly $Z_{cs}$ state to be  a $D^{*0}D_s^{-}/D^{0}D_s^{*-}/D^{*0}D_s^{*-}$ resonance with $I(J^P)=1/2(1^+, 0^-, 1^-, 2^-)$.
In Ref. \cite{Sun:2020hjw}, the  $Z_{cs}$ state in the molecular $ D_s^{*-} D^{*0}$  picture was studied and the obtained results supported that this state can be explained as a $ D_s^{*-} D^{*0}$  molecular resonance.
In Ref. \cite{Wang:2020rcx}, the mass spectra of the charmed strange tetraquark states with different quantum numbers in the molecular and diquark-antidiquark pictures were calculated  with the help of QCD sum rule method. The obtained results of both pictures are consistent with the experimental data.
In Ref. \cite{Wang:2020htx}, they studied the newly observed charmoniumlike state $Z_{cs}$ in the framework of chiral effective field theory up to the next-to-leading order with the explicit chiral dynamics. Their studies strongly supported that the $Z_{cs}$  is the partner of the $Z_{c}(3900)$ in the $SU(3)_f$ symmetry and the charmed strange molecular resonance with the same dynamical origin as the other charged heavy quarkonium like states.
In Ref. \cite{Wang:2020iqt}, the mass of $Z_{cs}$  state has been extracted in the diquark-antidiquark picture in the framework of the QCD sum rules and obtained results are consistent with the experimental data.
In Ref. \cite{Jin:2020yjn}, they investigated the strange hidden-charm tetraquark systems in the molecular and diquark-antidiquark pictures with the help of the chiral quark model and the quark delocalization colour screening model. They obtained same results in both quark models and they excluded the $D_{s}D^{*}/D_{s}^{*}D/D_{s}^{*}D^{*}$ molecular pictures for the $Z_{cs}$ state.
In Ref. \cite{Ikeno:2021ptx}, they have  studied the interaction of $\bar D_s D^*$ with the coupled channels $J/\psi K^-$, $K^{*-} \eta_c$, $D_s^- D^{*0}$, $D_s^{*-} D^0$ in the framework of the local hidden gauge approach. They have obtained that  the $D_s^- D^{*0}+ D_s^{*-} D^0$ combination couples to $J/\psi K^-$ and  $K^{*-} \eta_c$, but the $D_s^- D^{*0}- D_s^{*-} D^0$ combination does not. Moreover, they claimed that $D_s^- D^{*0}+ D_s^{*-} D^0$ interaction is not strong enough to produce a bound state or resonance, it is sufficient to produce a large accumulation of strength at the $\bar D_s D^*$ threshold in the $e^+ e^- \rightarrow K^+ ( D_s^{*-} D^0 + D_s^- D^{*0})$ reaction in agreement with experiment.
In Ref.  \cite{,Xu:2020evn}, the mass, residue and  magnetic moment of the $Z_{cs}$ state were investigated in the framework of the QCD sum rule and its extension in the weak electromagnetic field by using a molecular type interpolating current. The result of mass is consistent with the experiment.
As can be seen from these studies, the results obtained using different approaches  lead to  different interpretations for the  $Z_{cs}$ state.  More theoretical investigations are needed: Especially the strong decays of these states with the aim of determination of its width can be useful.  Future experimental results on different parameters of this state and their comparision with the theoretical predictions can help us understand the nature of this state.

\section{Discussion and concluding remarks}

Besides the spectroscopic parameters, the electromagnetic form factors and multipole moments of hadrons are important quantities  that carry information about the nature and quark-gluon organization of hadrons. The $ Z_ {cs} $  state is the first hidden-charmed tetraquark candidate composed of a strange quark. Based on the information provided by the experiment, we considered it composed of $ c \bar c s\bar u $ quarks/antiquarks with quantum numbers $J^P = 1^{+}$. We calculated the magnetic dipole moment of $ Z_ {cs} $  state both in the compact tetraquark of diquark-antidiquark and $D_s^{-}D^{*0}  $ / $ D_s^{*-} D^0 $ molecular pictures in LCSR using the on-shell photon DAs. We observed that the two pictures give close results, which are consistent with each other within the presented errors.   The obtained results in both pictures, are considerably large compared to the result of Ref.  \cite{,Xu:2020evn}, $ \mu_{Z_{cs}}=0.174 \pm 0.015~\mu_N$. This can be attributed to the fact that in Ref.  \cite{,Xu:2020evn},  the authors use the  extension of the QCD sum rule in the weak electromagnetic field approximation, while we apply  the full LCSR including all the photon's DAs without any approximation. 
Our results show reasonable  $ SU(3)_f $  violations of central values with the magnetic dipole moment of  $ Z_ {c}(3900) $, $ \mu_{Z_{c}}=0.67 \pm 0.32~\mu_N$, obtained in Ref.   \cite{Ozdem:2017jqh} using the compact tetraquark structure.

As it is clear, the two compact tetraquark of diquark-antidiquark and molecule pictures lead to roughly the same results for the magnetic dipole moments of $ Z_{cs} $. 
The existing theoretical predictions on the mass of this state and their comparison with the experimental value have also led to different assignments on the substructure of this state discussed above. More theoretical studies, especially on the strong decays of this state with the aim of prediction of its width may be very useful as the width of this state is also available from the experiment. Calculations of different parameters related to various interactions/decays of $ Z_{cs} $  state and their comparison with probable future experimental results can help us fix the quark-gluon organization and quantum numbers of this state. 


\appendix


\section*{Appendix: Photon Distribution Amplitudes and Wave Functions}

In this Appendix, we present the matrix elements $\langle \gamma(q)\vel \bar{q}(x) \Gamma_i q(0) \ver 0\rangle$  
and $\langle \gamma(q)\vel \bar{q}(x) \Gamma_i G_{\mu\nu}q(0) \ver 0\rangle$ in terms of  the photon DAs  and wave functions of different  twists \cite{Ball:2002ps},

\begin{eqnarray*}
\label{esbs14}
&&\langle \gamma(q) \vert  \bar q(x) \gamma_\mu q(0) \vert 0 \rangle
= e_q f_{3 \gamma} \left(\varepsilon_\mu - q_\mu \frac{\varepsilon
x}{q x} \right) \int_0^1 du e^{i \bar u q x} \psi^v(u)
\nonumber \\
&&\langle \gamma(q) \vert \bar q(x) \gamma_\mu \gamma_5 q(0) \vert 0
\rangle  = - \frac{1}{4} e_q f_{3 \gamma} \epsilon_{\mu \nu \alpha
\beta } \varepsilon^\nu q^\alpha x^\beta \int_0^1 du e^{i \bar u q
x} \psi^a(u)
\nonumber \\
&&\langle \gamma(q) \vert  \bar q(x) \sigma_{\mu \nu} q(0) \vert  0
\rangle  = -i e_q \langle \bar q q \rangle (\varepsilon_\mu q_\nu - \varepsilon_\nu
q_\mu) \int_0^1 du e^{i \bar u qx} \left(\chi \varphi_\gamma(u) +
\frac{x^2}{16} \mathbb{A}  (u) \right) \nonumber \\ 
&&-\frac{i}{2(qx)}  e_q \bar qq \left[x_\nu \left(\varepsilon_\mu - q_\mu
\frac{\varepsilon x}{qx}\right) - x_\mu \left(\varepsilon_\nu -
q_\nu \frac{\varepsilon x}{q x}\right) \right] \int_0^1 du e^{i \bar
u q x} h_\gamma(u)
\nonumber \\
&&\langle \gamma(q) | \bar q(x) g_s G_{\mu \nu} (v x) q(0) \vert 0
\rangle = -i e_q \langle \bar q q \rangle \left(\varepsilon_\mu q_\nu - \varepsilon_\nu
q_\mu \right) \int {\cal D}\alpha_i e^{i (\alpha_{\bar q} + v
\alpha_g) q x} {\cal S}(\alpha_i)
\nonumber \\
&&\langle \gamma(q) | \bar q(x) g_s \tilde G_{\mu \nu}(v
x) i \gamma_5  q(0) \vert 0 \rangle = -i e_q \langle \bar q q \rangle \left(\varepsilon_\mu q_\nu -
\varepsilon_\nu q_\mu \right) \int {\cal D}\alpha_i e^{i
(\alpha_{\bar q} + v \alpha_g) q x} \tilde {\cal S}(\alpha_i)
\nonumber \\
&&\langle \gamma(q) \vert \bar q(x) g_s \tilde G_{\mu \nu}(v x)
\gamma_\alpha \gamma_5 q(0) \vert 0 \rangle = e_q f_{3 \gamma}
q_\alpha (\varepsilon_\mu q_\nu - \varepsilon_\nu q_\mu) \int {\cal
D}\alpha_i e^{i (\alpha_{\bar q} + v \alpha_g) q x} {\cal
A}(\alpha_i)
\nonumber \\
&&\langle \gamma(q) \vert \bar q(x) g_s G_{\mu \nu}(v x) i
\gamma_\alpha q(0) \vert 0 \rangle = e_q f_{3 \gamma} q_\alpha
(\varepsilon_\mu q_\nu - \varepsilon_\nu q_\mu) \int {\cal
D}\alpha_i e^{i (\alpha_{\bar q} + v \alpha_g) q x} {\cal
V}(\alpha_i) \nonumber\\
&& \langle \gamma(q) \vert \bar q(x)
\sigma_{\alpha \beta} g_s G_{\mu \nu}(v x) q(0) \vert 0 \rangle  =
e_q \langle \bar q q \rangle \left\{
        \left[\left(\varepsilon_\mu - q_\mu \frac{\varepsilon x}{q x}\right)\left(g_{\alpha \nu} -
        \frac{1}{qx} (q_\alpha x_\nu + q_\nu x_\alpha)\right) \right. \right. q_\beta
\nonumber \\ && -
         \left(\varepsilon_\mu - q_\mu \frac{\varepsilon x}{q x}\right)\left(g_{\beta \nu} -
        \frac{1}{qx} (q_\beta x_\nu + q_\nu x_\beta)\right) q_\alpha
-
         \left(\varepsilon_\nu - q_\nu \frac{\varepsilon x}{q x}\right)\left(g_{\alpha \mu} -
        \frac{1}{qx} (q_\alpha x_\mu + q_\mu x_\alpha)\right) q_\beta
\nonumber \\ &&+
         \left. \left(\varepsilon_\nu - q_\nu \frac{\varepsilon x}{q.x}\right)\left( g_{\beta \mu} -
        \frac{1}{qx} (q_\beta x_\mu + q_\mu x_\beta)\right) q_\alpha \right]
   \int {\cal D}\alpha_i e^{i (\alpha_{\bar q} + v \alpha_g) qx} {\cal T}_1(\alpha_i)
\nonumber \\ &&+
        \left[\left(\varepsilon_\alpha - q_\alpha \frac{\varepsilon x}{qx}\right)
        \left(g_{\mu \beta} - \frac{1}{qx}(q_\mu x_\beta + q_\beta x_\mu)\right) \right. q_\nu
\nonumber \\ &&-
         \left(\varepsilon_\alpha - q_\alpha \frac{\varepsilon x}{qx}\right)
        \left(g_{\nu \beta} - \frac{1}{qx}(q_\nu x_\beta + q_\beta x_\nu)\right)  q_\mu
\nonumber \\ && -
         \left(\varepsilon_\beta - q_\beta \frac{\varepsilon x}{qx}\right)
        \left(g_{\mu \alpha} - \frac{1}{qx}(q_\mu x_\alpha + q_\alpha x_\mu)\right) q_\nu
\nonumber \\ &&+
         \left. \left(\varepsilon_\beta - q_\beta \frac{\varepsilon x}{qx}\right)
        \left(g_{\nu \alpha} - \frac{1}{qx}(q_\nu x_\alpha + q_\alpha x_\nu) \right) q_\mu
        \right]      
    \int {\cal D} \alpha_i e^{i (\alpha_{\bar q} + v \alpha_g) qx} {\cal T}_2(\alpha_i)
\nonumber \\
&&+\frac{1}{qx} (q_\mu x_\nu - q_\nu x_\mu)
        (\varepsilon_\alpha q_\beta - \varepsilon_\beta q_\alpha)
    \int {\cal D} \alpha_i e^{i (\alpha_{\bar q} + v \alpha_g) qx} {\cal T}_3(\alpha_i)
\nonumber \\ &&+
        \left. \frac{1}{qx} (q_\alpha x_\beta - q_\beta x_\alpha)
        (\varepsilon_\mu q_\nu - \varepsilon_\nu q_\mu)
    \int {\cal D} \alpha_i e^{i (\alpha_{\bar q} + v \alpha_g) qx} {\cal T}_4(\alpha_i)
                        \right\}~,
\end{eqnarray*}
where $\varphi_\gamma(u)$ is the distribution amplitude of leading twist-2, $\psi^v(u)$,
$\psi^a(u)$, ${\cal A}(\alpha_i)$ and ${\cal V}(\alpha_i)$, are the twist-3 amplitudes, and
$h_\gamma(u)$, $\mathbb{A}(u)$, ${\cal S}(\alpha_i)$, ${\cal{\tilde S}}(\alpha_i)$, ${\cal T}_1(\alpha_i)$, ${\cal T}_2(\alpha_i)$, ${\cal T}_3(\alpha_i)$ 
and ${\cal T}_4(\alpha_i)$ are the
twist-4 photon DAs.
The measure ${\cal D} \alpha_i$ is defined as
\begin{eqnarray*}
\label{nolabel05}
\int {\cal D} \alpha_i = \int_0^1 d \alpha_{\bar q} \int_0^1 d
\alpha_q \int_0^1 d \alpha_g \delta(1-\alpha_{\bar
q}-\alpha_q-\alpha_g)~.\nonumber
\end{eqnarray*}

The expressions of the DAs entering into the above matrix elements are
defined as:

\begin{eqnarray}
\varphi_\gamma(u) &=& 6 u \bar u \left( 1 + \varphi_2(\mu)
C_2^{\frac{3}{2}}(u - \bar u) \right),
\nonumber \\
\psi^v(u) &=& 3 \left(3 (2 u - 1)^2 -1 \right)+\frac{3}{64} \left(15
w^V_\gamma - 5 w^A_\gamma\right)
                        \left(3 - 30 (2 u - 1)^2 + 35 (2 u -1)^4
                        \right),
\nonumber \\
\psi^a(u) &=& \left(1- (2 u -1)^2\right)\left(5 (2 u -1)^2 -1\right)
\frac{5}{2}
    \left(1 + \frac{9}{16} w^V_\gamma - \frac{3}{16} w^A_\gamma
    \right),
\nonumber \\
h_\gamma(u) &=& - 10 \left(1 + 2 \kappa^+\right) C_2^{\frac{1}{2}}(u
- \bar u),
\nonumber \\
\mathbb{A}(u) &=& 40 u^2 \bar u^2 \left(3 \kappa - \kappa^+
+1\right)  +
        8 (\zeta_2^+ - 3 \zeta_2) \left[u \bar u (2 + 13 u \bar u) \right.
\nonumber \\ && + \left.
                2 u^3 (10 -15 u + 6 u^2) \ln(u) + 2 \bar u^3 (10 - 15 \bar u + 6 \bar u^2)
        \ln(\bar u) \right],
\nonumber \\
{\cal A}(\alpha_i) &=& 360 \alpha_q \alpha_{\bar q} \alpha_g^2
        \left(1 + w^A_\gamma \frac{1}{2} (7 \alpha_g - 3)\right),
\nonumber \\
{\cal V}(\alpha_i) &=& 540 w^V_\gamma (\alpha_q - \alpha_{\bar q})
\alpha_q \alpha_{\bar q}
                \alpha_g^2,
\nonumber \\
{\cal T}_1(\alpha_i) &=& -120 (3 \zeta_2 + \zeta_2^+)(\alpha_{\bar
q} - \alpha_q)
        \alpha_{\bar q} \alpha_q \alpha_g,
\nonumber \\
{\cal T}_2(\alpha_i) &=& 30 \alpha_g^2 (\alpha_{\bar q} - \alpha_q)
    \left((\kappa - \kappa^+) + (\zeta_1 - \zeta_1^+)(1 - 2\alpha_g) +
    \zeta_2 (3 - 4 \alpha_g)\right),
\nonumber \\
{\cal T}_3(\alpha_i) &=& - 120 (3 \zeta_2 - \zeta_2^+)(\alpha_{\bar
q} -\alpha_q)
        \alpha_{\bar q} \alpha_q \alpha_g,
\nonumber \\
{\cal T}_4(\alpha_i) &=& 30 \alpha_g^2 (\alpha_{\bar q} - \alpha_q)
    \left((\kappa + \kappa^+) + (\zeta_1 + \zeta_1^+)(1 - 2\alpha_g) +
    \zeta_2 (3 - 4 \alpha_g)\right),\nonumber \\
{\cal S}(\alpha_i) &=& 30\alpha_g^2\{(\kappa +
\kappa^+)(1-\alpha_g)+(\zeta_1 + \zeta_1^+)(1 - \alpha_g)(1 -
2\alpha_g)\nonumber +\zeta_2[3 (\alpha_{\bar q} - \alpha_q)^2-\alpha_g(1 - \alpha_g)]\},\nonumber \\
\tilde {\cal S}(\alpha_i) &=&-30\alpha_g^2\{(\kappa -\kappa^+)(1-\alpha_g)+(\zeta_1 - \zeta_1^+)(1 - \alpha_g)(1 -
2\alpha_g)\nonumber +\zeta_2 [3 (\alpha_{\bar q} -\alpha_q)^2-\alpha_g(1 - \alpha_g)]\}.
\end{eqnarray}

Numerical values of parameters used in distribution amplitudes are: $\varphi_2(1~GeV) = 0$, 
$w^V_\gamma = 3.8 \pm 1.8$, $w^A_\gamma = -2.1 \pm 1.0$, 
$\kappa = 0.2$, $\kappa^+ = 0$, $\zeta_1 = 0.4$, $\zeta_2 = 0.3$.


\bibliography{article}

\end{document}